\documentclass[twocolumn]{aastex631}
\usepackage{CJK}
\usepackage{mathtools}

\begin{document}
\begin{CJK*}{UTF8}{gbsn}

\title{Solar Wind Structures from the Gaussianity of Magnetic Magnitude}

\author[0000-0001-9570-5975]{Zesen Huang*}
\affiliation{Department of Earth, Planetary, and Space Sciences, University of California, Los Angeles, CA, USA}
\author[0000-0002-2582-7085]{Chen Shi*}
\affiliation{Department of Earth, Planetary, and Space Sciences, University of California, Los Angeles, CA, USA}
\author[0000-0002-2381-3106]{Marco Velli*}
\affiliation{Department of Earth, Planetary, and Space Sciences, University of California, Los Angeles, CA, USA}
\author[0000-0002-1128-9685]{Nikos Sioulas}
\affiliation{Department of Earth, Planetary, and Space Sciences, University of California, Los Angeles, CA, USA}
\author[0000-0002-4440-7166]{Olga Panasenco}
\affiliation{Advanced Heliophysics, Pasadena, CA 91106, USA}
\author[0000-0002-4625-3332]{Trevor Bowen}
\affiliation{Space Sciences Laboratory, University of California, Berkeley, CA 94720-7450, USA}
\author[0000-0002-6276-7771]{Lorenzo Matteini}
\affiliation{Imperial College London, South Kensington Campus, London SW7 2AZ, UK}
\author[0000-0002-2116-4712]{Mingtao Xia}
\affiliation{Courant Institute of Mathematical Sciences, New York University, New York, NY 10012, United States of America}
\author[0000-0003-3367-5074]{Xiaofei Shi}
\affiliation{Department of Earth, Planetary, and Space Sciences, University of California, Los Angeles, CA, USA}
\author[0000-0002-9951-8787]{Sheng Huang}
\affiliation{Center for Space Physics, Boston University, Boston, MA, USA}
\author[0000-0002-9954-4707]{Jia Huang}
\affiliation{Space Sciences Laboratory, University of California, Berkeley, CA 94720-7450, USA}
\author[0000-0001-6248-183X]{Lizet Casillas}
\affiliation{Department of Earth, Planetary, and Space Sciences, University of California, Los Angeles, CA, USA}

\correspondingauthor{Zesen Huang, Marco Velli, Chen Shi}
\email{zesenhuang@ucla.edu; mvelli@ucla.edu; cshi1993@ucla.edu}


\begin{abstract}
One of the primary science objectives of Parker Solar Probe (PSP) is to determine the structures and dynamics of the plasma and magnetic fields at the sources of the solar wind. However, establishing the connection between {\it in situ} measurements and structures and dynamics in the solar atmosphere is challenging: most of the magnetic footpoint mapping techniques have significant uncertainties in the source localization of a plasma parcel observed {\it in situ}, and the PSP plasma measurements suffer from a limited field of view. Therefore it is of interest to investigate whether {\it in situ} measurements can be used on their own to identify streams originating from the same structures in the corona more finely than the well known fast wind-coronal hole, slow wind-elsewhere distinction. Here we develop a novel time series visualization method \textcolor{red}{(time-frequency representation or TFR)} named Gaussianity Scalogram. Utilizing this method, by analyzing the magnetic magnitude data from both PSP and Ulysses, we successfully identify {\it in situ} structures that are possible remnants of solar atmospheric and magnetic structures spanning more than seven orders of magnitude, from years to seconds, including polar and mid-latitude coronal holes, as well as structures compatible with super-granulation , ``jetlets'' and ``picoflares''. \textcolor{red}{Furthermore, computer simulations of Alfv\'enic turbulence successfully reproduce the Gaussianization of the magnetic magnitude for locally homogeneous structures.} Building upon these discoveries, the Gaussianity Scalogram can help future studies to reveal the fractal-like fine structures in the solar wind time series from both PSP and decades-old data archive.

\end{abstract}


\section{Introduction}

The solar atmosphere is highly structured both spatially and temporally \citep{bale_highly_2019,kasper_alfvenic_2019}. Recent studies have successfully established connections between PSP (\cite{fox_solar_2016} and see \cite{raouafi_parker_2023} for a review for results of the first four years of the mission) {\it in situ} observations and solar atmospheric structures including mid-latitude coronal holes \citep{badman_prediction_2023,davis_evolution_2023}, pseudostreamers \citep{kasper_parker_2021}, and supergranulation \citep{bale_solar_2021, fargette_characteristic_2021,bale_interchange_2023}, even though alternative explanations remain \citep{shi_patches_2022}. Recent advances in remote sensing provide strong support for the minutes long small-scale jetting activity from magnetic reconnection (``jetlets'') as a major source of the solar wind \citep{raouafi_magnetic_2023}. In addition, EUV observations from Solar Orbiter \citep{muller_solar_2020} unveiled ubiquitous brightening termed ``picoflare'' \citep{chitta_picoflare_2023} with associated jets that last only a few tens of seconds, suggesting the solar wind source might be highly intermittent. However, magnetic footpoint mapping methods \citep{badman_magnetic_2020,panasenco_exploring_2020,badman_prediction_2023} use photospheric magnetic field observations over the whole visible disk that are refreshed at best once every six hours and lack, of course, any real temporal reliability for the far side. Therefore, such methods are hardly able to reliably contextualize and explain the boundaries of the highly structured solar wind {\it in situ} time series, except perhaps in a statistical sense.

\textcolor{red}{Previous studies on the distribution of solar wind parameters primarily focused on the statistical properties of the solar wind magnetic field and plasma moments at 1 AU and beyond. For example, \cite{whang_probability_1977} and \cite{padhye_distribution_2001} found that the magnetic field components follow near-Gaussian distributions, and \cite{bandyopadhyay_statistics_2020} showed similar properties in Earth's magnetosheath turbulence. For the magnetic field magnitude, \cite{burlaga_lognormal_2000} and \cite{burlaga_lognormal_2001} demonstrated that it generally follows a log-normal distribution. On the simulation side, \cite{yamamoto_gaussian_1991} conducted a 3D hydrodynamic simulation and found that the velocity components follow a Gaussian distribution at small wave numbers and a long-tail distribution at larger wave numbers. It is hence empirically expected that the magnetic field components $ B_{x,y,z} $ follow a Gaussian distribution, and thus the magnetic magnitude $ B=\sqrt{B_x^2+B_y^2+B_z^2} $ follows a $\chi$ or Maxwell-Boltzmann distribution. However, for the solar wind measured at PSP perihelion distances, the distributions of magnetic field components generally deviate significantly from Gaussian due to the presence of large amplitude spherically polarized Alfv\'en waves (i.e., switchbacks, see e.g., \cite{bale_highly_2019,larosa_switchbacks_2021,drake_switchbacks_2021,dudok_de_wit_switchbacks_2020,shi_patches_2022,bale_interchange_2023}). Therefore, it is not expected from any of the previous studies that the magnetic magnitude should follow a Gaussian distribution or any known distributions.}

\textcolor{red}{Contrary to previous attempts, using a data mining approach, this study starts by reporting an unexpected property of the solar wind magnetic field. We find that the magnetic magnitude $B$, once normalized by a power law fit with regard to the heliocentric distance, occasionally exhibits perfect Gaussian distributions. Based on this discovery, we introduce a novel time series visualization method (time-frequency representation or TFR) named Gaussianity Scalogram (GS) to visualize the spatial-scale (time-frequency) dependent Gaussianity of $ B $. Applying this method to data from PSP and Ulysses, we find that the Gaussian intervals successfully map to the \textit{in situ} remnants of coronal holes \citep{badman_prediction_2023,davis_evolution_2023} and switchback patches \citep{bale_solar_2021,shi_patches_2022,bale_interchange_2023}. On smaller scales, Gaussian intervals map to structures that are temporally compatible with small-scale jetting activity, including "jetlets" and "picoflares" \citep{raouafi_magnetic_2023,chitta_picoflare_2023}. }The rest of the paper is structured as follows: In the next section, we describe the helio-radial power law fit and the construction of GS; in Section 3, we present some applications of GS; in Section 4, we discuss the computer simulation of Gaussianization of $ B $ in MHD turbulence and the major implications; in Section 5, we conclude and summarize our results.

\section{Helioradial Dependence of $B$ and the Gaussianity Scalogram}

\begin{figure*}[ht]
    \includegraphics[width=.90\textwidth]{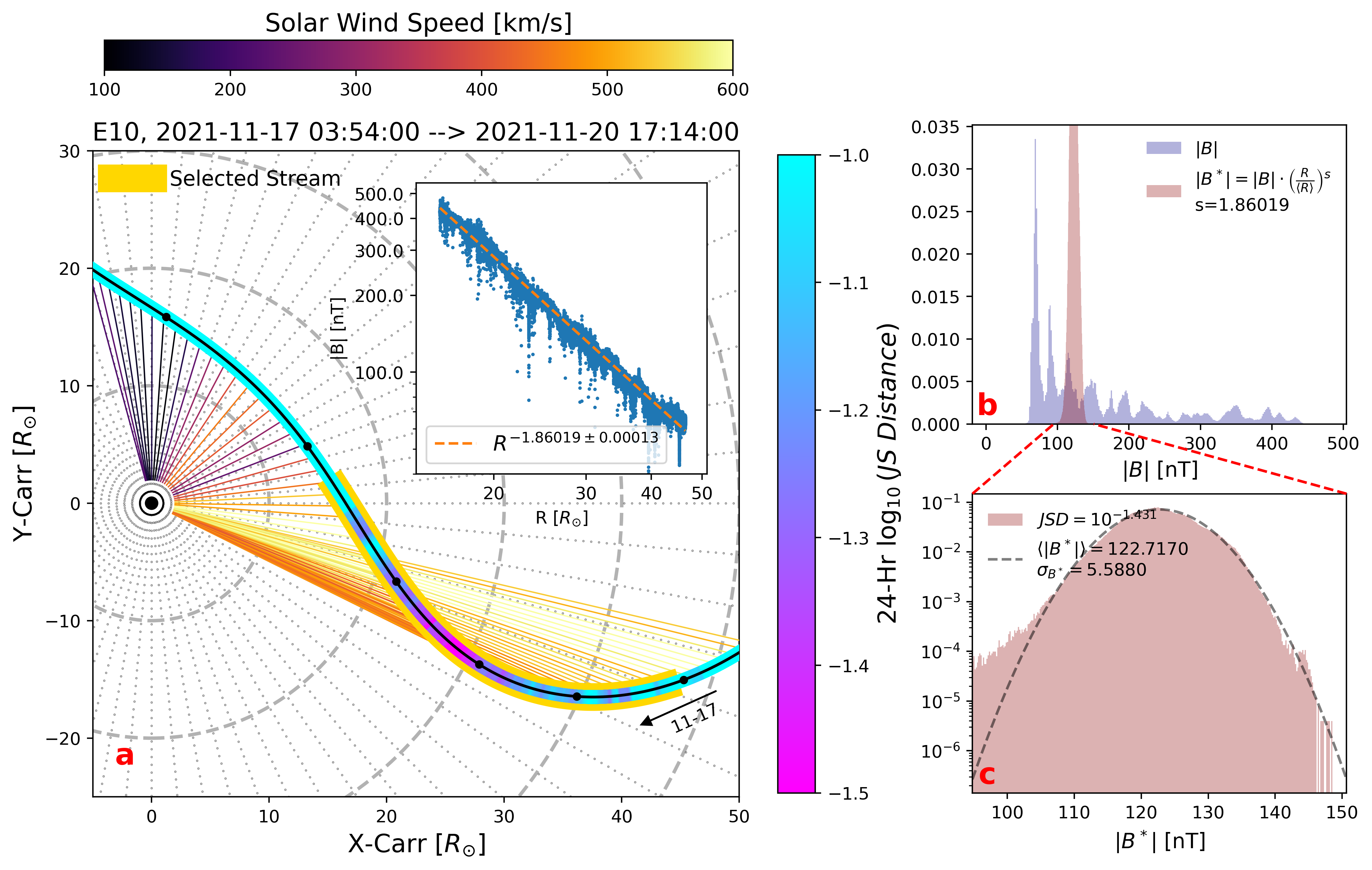}
    \includegraphics[width=.95\textwidth]{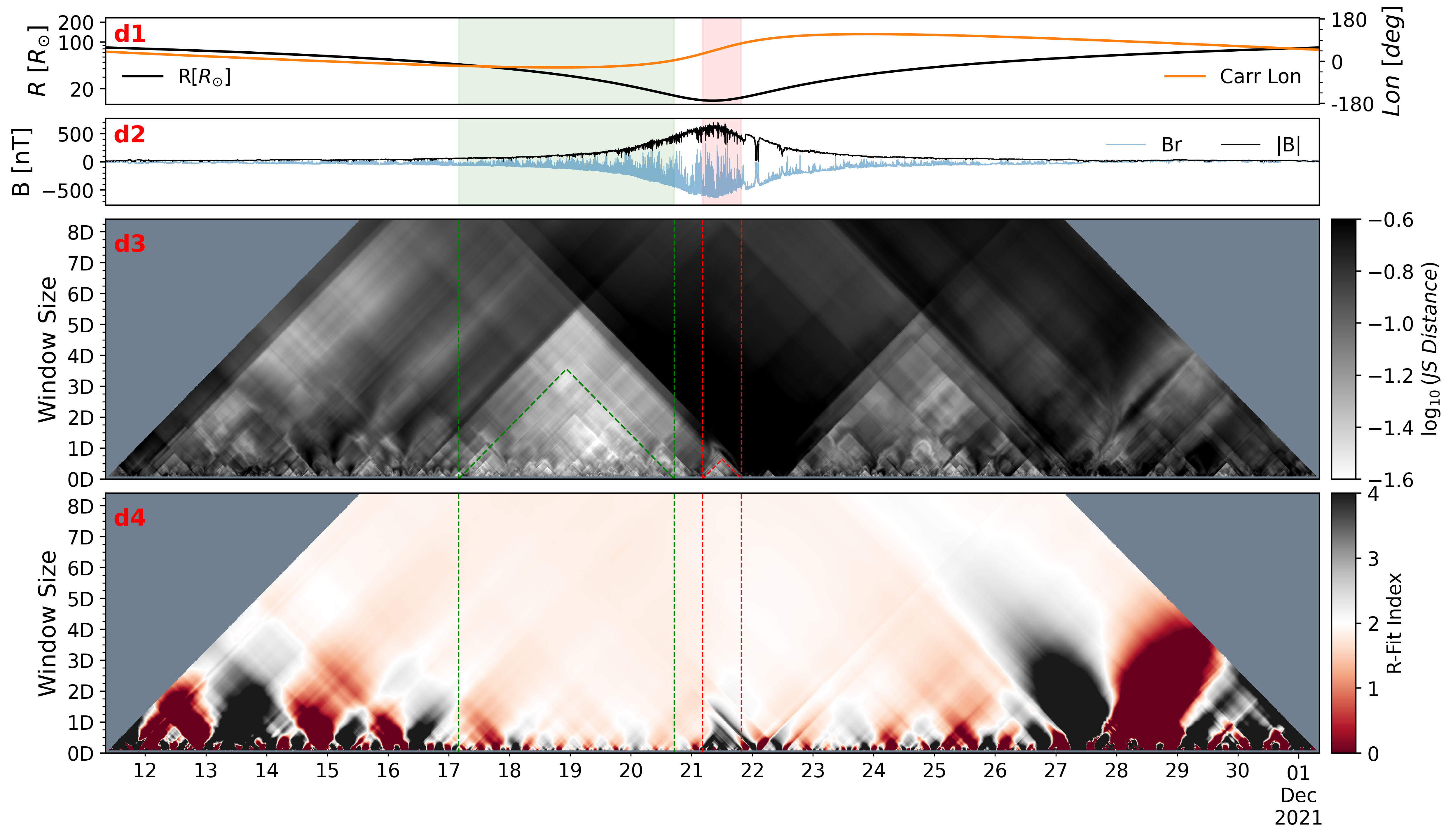}
    \caption{Selected interval from November 2021, encounter 10 of Parker Solar Probe. (a): This panel presents the spacecraft trajectory in the Carrington corotating frame from the afternoon of November 16, 2021, to the afternoon of November 22, 2021. Each day's start is indicated with black circles. The ballistic solar wind streamlines are plotted at a 2-hour cadence and colored according to the 10-minute averaged solar wind speed profile from SPAN-ion moment. The selected interval is emphasized with a golden bar, and the 24-Hour window Jensen-Shannon Distance (JSD) of normalized magnetic magnitude $B^*$ is represented by the colored band. An inset displays the helio-radial power law dependence of $B$. (b): The histogram of $B$ and $B^*$ from the selected interval. (c): The histogram of $B^*$ and $JSD(PDF_{B^*})$. (d1): Spacecraft heliocentric distance (black) and Carrington longitude (orange). (d2): Magnetic field radial component $B_r$ and magnitude $B$. (d3):Gaussianity Scalogram (Scalogram of JSD). The selected interval is highlighted with the green pyramid. (d4): Helio-radial power law fit index scalogram of $B$.}
    \label{fig:E10}
\end{figure*}

Two of the most interesting yet overlooked features of the time series of the solar wind magnetic field magnitude $B$ are that: 1. Sometimes $B$ displays a surprisingly stable power law dependence on the heliocentric distance $R$; 2. By applying a helio-radial power law fit between $B$ and $R$, i.e. $B\propto R^{-s}$, the fit normalized magnetic magnitude $B^*=B (R/R_0)^s$ sometimes displays a near-perfect Gaussian distribution. This is illustrated in Figure \ref{fig:E10} (a-c), where the selected interval is highlighted with a golden bar in panel (a) and the helio-radial power law fit (fit index $s=1.86$) is shown in the inset figure. The histogram of $B$ is shown in blue in panel (b) and the normalized $B^*$ is shown in red. To illustrate the close proximity of the probability density function of $B^*$ ($PDF_{B^*}$) to a Gaussian distribution ($\mathcal{N}$), a standard Gaussian curve is overplotted in panel (c) (shifted with the mean value $\langle B^*\rangle$ and scaled with the standard deviation $\sigma_{B^*}$). The Jensen-Shannon Distance (JSD) of base $e$ (a statistical distance metric between probability density functions \citep{lin_divergence_1991}) is calculated between $PDF_{B^*}$ and $\mathcal{N}$, and the value is $JSD(PDF_{B^*},\mathcal{N})=10^{-1.431}$, indicating considerable closeness between two distribution functions (for benchmark, see appendix). In addition, this highly Gaussian $B^*$ interval coincides with the radial solar wind speed profile which is visualized with radial colored lines in panel (a) and Figure \ref{fig:panorama} (c) (compiled with SPAN-ion from SWEAP suite \cite{kasper_solar_2016}). From Nov-17 to Nov-20, the spacecraft was immersed in the high speed solar wind, indicating its coronal hole origin. The JSD produced by this process is represented as one pixel (tip of the green pyramid) in the Gaussianity Scalogram (GS) shown in panel (d3), and the scalogram for the corresponding helio-radial power law fit index $s$ is displayed in panel (d4).


Each pixel in the GS is characterized by a timestamp ($t_{mid}$) and window size ($win$), similar to wavelet scalogram. Uniquely in GS, the step size in $win$ (vertical axis) is chosen to be twice the step size in $t_{mid}$ (horizontal axis), and thus the time range covered by one pixel corresponds to the same time range covered by three pixels in the following row, and so on towards the smallest scales. Therefore, if an interval and the nested sub-intervals possess similar characteristics (e.g. relatively small JSD regardless of $t_{mid}$ and $win$ within the interval), a pyramidal structure is expected from the GS, and the base of the pyramid indicates the start and end time of the interval. One example is highlighted by the green dashed pyramid in panel (d3), where the tip of the pyramid is in fact selected {\it a posteriori} as the local minimum in the GS ($PDF_{B^*}$ being closest to Gaussian among the surrounding time and scales). Ample information can be inferred from the GS: 1. A semi-crossing of the heliospheric current sheet (HCS) at noon of Nov-22 is visualized as an inverted black pyramid. This is because the rapid drop of $B$ during HCS crossing can significantly destroy its Gaussianity; 2. It has been confirmed recently by \cite{badman_prediction_2023} that the solar wind can be traced back to a single mid-latitude coronal hole from Nov-17 to the end of Nov-20, and from another coronal hole for the whole day of Nov-21 (see also \cite{bale_interchange_2023,panasenco_exploring_2020,badman_magnetic_2020}). The coronal holes are naturally visualized here as two white pyramids (green and red dashed lines) separated by a dark region around the mid-night of Nov-20; 3. The helio-radial power law fit index $s$ is unexpectedly stable regardless of locations and scales and systematically deviates from $R^{-2}$ ($s\simeq 1.87 \pm 0.02$) (see also \cite{bale_highly_2019}).

\begin{figure*}[ht]
    \includegraphics[width=.70\textwidth]{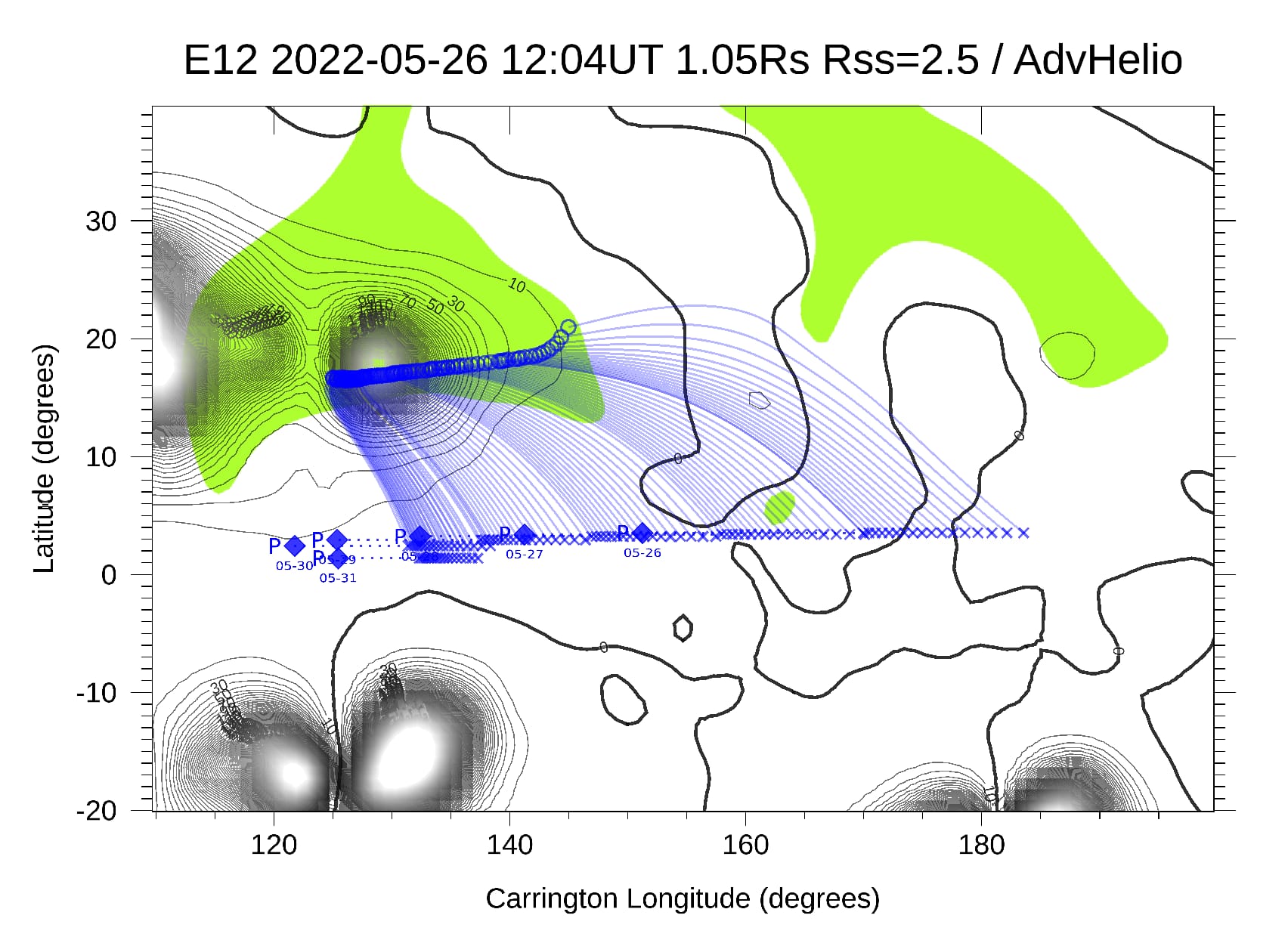}
    \caption{Magnetic field connectivity with the solar sources during inbound of PSP encounter 12. The thick black lines are the model neutral lines. Black contours indicate magnetic field pressure at 1.05 $R_s$. The ballistic projection of the PSP trajectory (blue diamonds) on the source surface (blue crosses) and down to the solar wind source regions (blue circles) is calculated for source surfaces $R_{ss}/R_s$ = 2.5 (see \cite{panasenco_exploring_2020} for details) and measured in situ solar wind speed ± 80 km s−1. Open magnetic field regions are shown in blue (negative) and green (positive).}
    \label{fig:PFSS_E12}
\end{figure*}

The clear correspondence between the white pyramid and coronal hole encourages us to predict intervals of solar wind originating from coronal holes with GS compiled from PSP data. Among the first 14 encounters (Nov-2018 to Dec-2022), we only identified one more (for a total of 2) long intervals ($>$ 3 days) characterized by high Gaussianity in $B^*$. A panoramic view of these two long intervals is shown in Figure \ref{fig:panorama}. The newly found interval from the inbound of E12, shown in Figure \ref{fig:E12} and Figure \ref{fig:panorama} (d), is characterized by a 5-day long highly Gaussian $B^*$ time series. For illustration purpose, the green pyramid in Figure \ref{fig:E12} (d3) is selected as the deepest local minimum in GS for $win >$ 3 days. The histogram of $B^*$ is remarkably concentrated (panel (b)) and aligns with Gaussian almost perfectly within 4 standard deviation (panel (c)). Similar to Figure \ref{fig:E10} (c), the non-Gaussian part of $PDF_{B^*}$ has a systematic bias towards magnetic holes (weaker magnetic magnitudes, for recent studies using PSP data see e.g. \cite{yu_characteristics_2021,yu_small-scale_2022}), and the helio-radial power law fit index scalogram also shows a systematic deviation from $s=2$, similarly $s\simeq 1.87 \pm 0.02$. To validate this prediction, independent results from Potential Field Source Surface (PFSS) modeling is shown in Figure \ref{fig:PFSS_E12} (see \cite{panasenco_exploring_2020} for more details), which indicates that the selected interval is indeed magnetically connected to a mid-latitude coronal hole.

\section{Fractal-Like Gaussian Structures in the Solar Wind}

To substantiate the applications of GS, here we demonstrate several examples that visualize the fractal-like Gaussian structures in the solar wind (Due to the rapid movement of PSP around perihelia, the structures in the {\it in situ} time series can be categorized into two kinds. Spatial: longitudinal structures traversed by PSP; Temporal (radial): radial structures advected by the solar wind and/or propagation of Alfv\'en waves) based on Taylor Hypothesis \citep{perez_applicability_2021}. From the largest scales: Ulysses, years-long polar coronal hole \citep{mccomas_three-dimensional_2003}, towards the smaller scales: hour-long switchback patches \citep{bale_solar_2021,fargette_characteristic_2021,shi_patches_2022,bale_interchange_2023}; minutes-long structures compatible with ``jetlets'' \cite{raouafi_magnetic_2023}; and second-long structures compatible with ``picoflare'' \cite{chitta_picoflare_2023}.

\begin{figure*}[ht]
    \includegraphics[width=.95\textwidth]{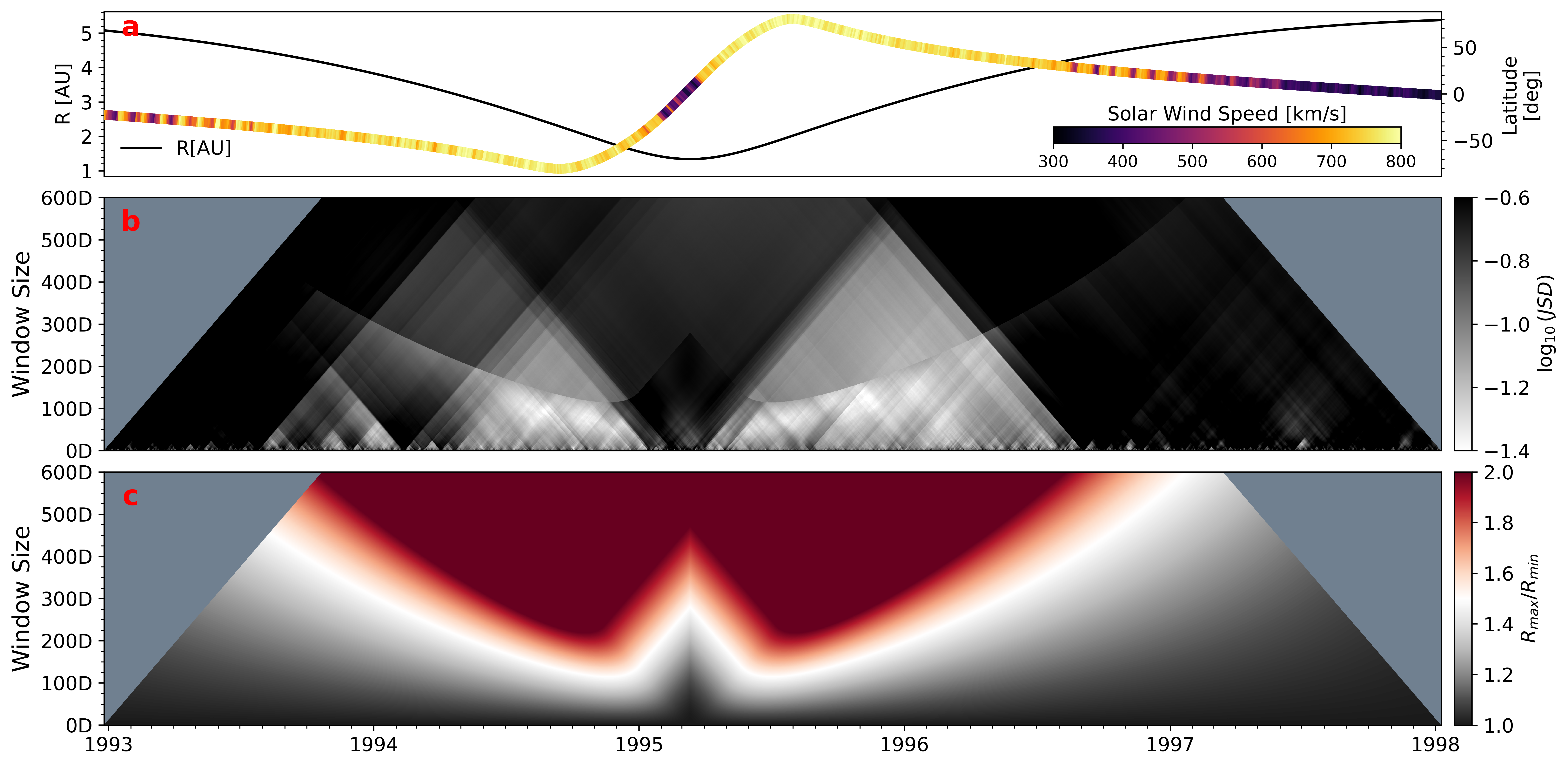}
    \caption{Gaussianity Scalogram from Ulysses first orbit from 1993 to 1998. (a): Ulysses heliocentric distance $R$ (black line) and heliographic latitude colored with local 48-hour averaged solar wind speed. (b): GS compiled from magnetic magnitude $B$ (lower half) and helio-radial power law normalized magnitude $B^*$ (upper half). (c): $R_{max}/R_{min}$ of each interval (pixel), the cut-off value is chosen to be $R_{max}/R_{min}=1.5$, beyond which $B$ is normalized into $B^*$ using helio-radial power law fit before calculating the Gaussianity.}
    \label{fig:ulysses}
\end{figure*}

Figure \ref{fig:ulysses} shows the GS of the first Ulysses orbit, and the colorbar in panel (b) is enhanced compared to Figure \ref{fig:E10} (d3) for illustration purposes. The solar latitude and wind speed profile in panel (a) indicate that the spacecraft was in southern and northern polar coronal holes in the whole year of 1994, and from 1995 to 1997 (see also \cite{mccomas_three-dimensional_2003}). The two large white pyramids in the GS clearly correspond to the two polar coronal holes. Notably, the boundary observed in panel (b) results from an artificial cut-off in the helio-radial power law fit, as shown in panel (c). The cut-off value is chosen to be $R_{max}/R_{min} = 1.5$, below which the spacecraft is considered stationary, and $B$ is not normalized to $B^*$ using helio-radial power law fit prior to the calculation of Gaussianity. However, the Gaussianity is much weaker in the polar coronal holes compared to the mid-latitude coronal holes observed by PSP at much smaller heliocentric distance, and the histograms of magnetic magnitude show much more significant fat tail towards the magnetic holes side (not shown here). This indicates that the Gaussianity of magnetic magnitudes decreases with increasing heliocentric distance, possibly due to the transition from low-$\beta$ to high-$\beta$ environment ($\beta=2\mu_0P/B^2$ is the ratio between plasma thermal pressure $P$ and magnetic pressure $B^2/2\mu_0$), and the plasma thermal pressure hence has a larger influence on the distribution of $B$. Additionally this also indicates that magnetic holes are much more preferred than spikes in the solar wind plasma.

\begin{figure*}[ht]
    \includegraphics[width=.95\textwidth]{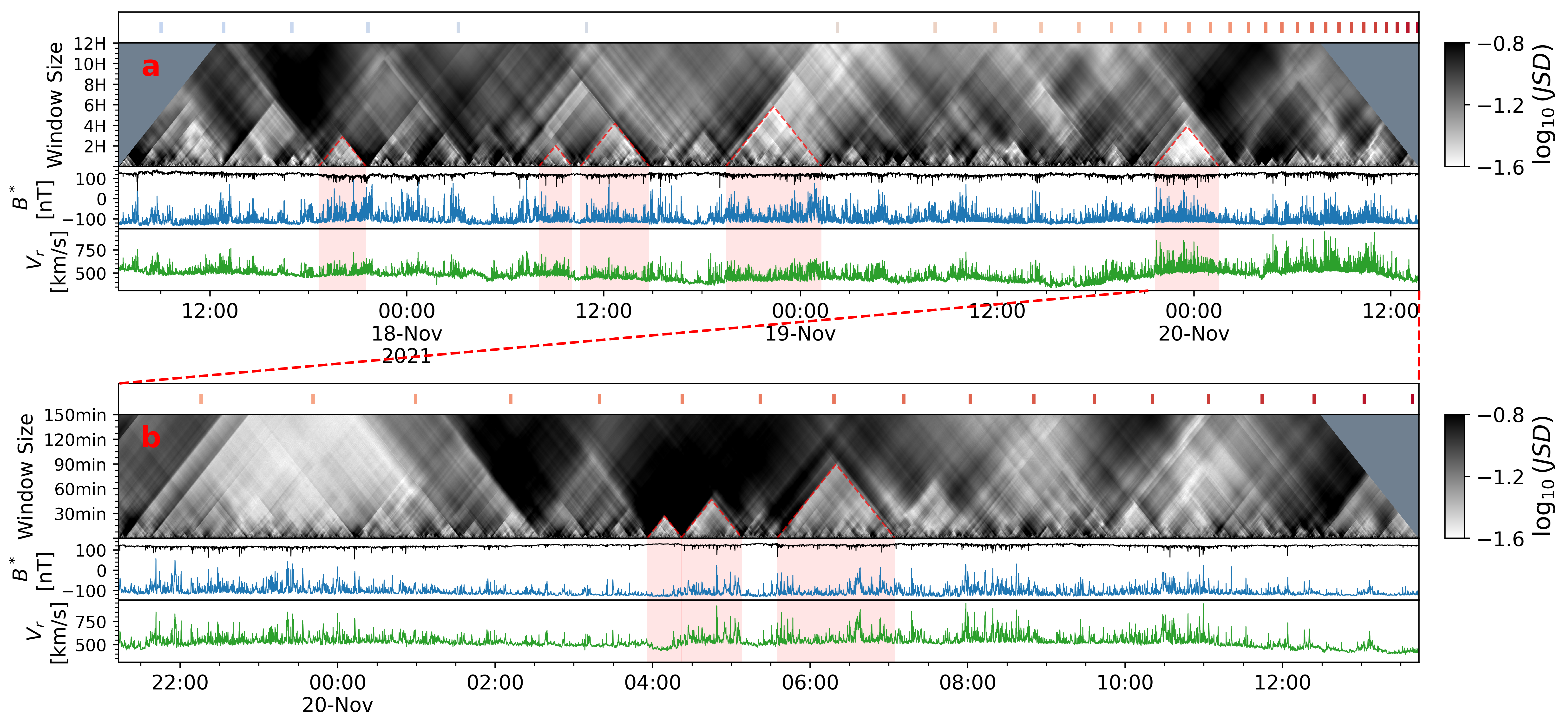}
    \caption{Gaussianity scalogram of Parker Solar Probe E10 inbound mid-latitude coronal hole. (a): From top to bottom: Carrington longitude plotted with 1 degree cadence colored with spacecraft angular velocity in the corotating frame (blue: prograde, red: retrograde); JS scalogram with 1-minute resolution; Normalized magnetic magnitude $B^*$ and radial component $B_r^*$; Radial solar wind speed $V_r$; (b): Expanded view of panel (a) with 10-second resolution.}
    \label{fig:switchback patches}
\end{figure*}

Figure \ref{fig:switchback patches} shows the hour-long switchback patches from a single mid-latitude coronal hole in PSP E10, which have been recently proposed to be the remnants of the supergranulations in the solar atmosphere \citep{bale_solar_2021,fargette_characteristic_2021,bale_interchange_2023}. To indicate the spacecraft movement, the Carrington longitude of PSP is plotted every one degree on the top bars of both panels (a) and (b), and the color indicates spacecraft angular velocity in the corotating frame (blue: prograde, red: retrograde, see also Figure \ref{fig:E10} for PSP trajectory in the corotating frame at the same perihelion). The magnetic magnitude is normalized with a universal helio-radial power law fit index ($s=1.87$) and the GS is compiled with the high-resolution fluxgate magnetic data ($\sim$ 292 Hz, see \cite{bale_fields_2016,bowen_merged_2020}). The red dashed pyramids in panel (a) and (b) are drawn to highlight the $B^*$ intervals with high level of Gaussianity. The selected intervals in panel (a) show that the GS effectively captures some of the switchback patches. When these are compared with the Carrington longitude, it becomes evident that some of the structures align with the size of supergranulation, as discussed in \cite{bale_interchange_2023}. However, other structures, which are smaller in angular size and likely temporal in nature, could be more accurately attributed to the 'breathing' phenomenon of the solar wind, as explained in \cite{berger_quiescent_2017,shi_patches_2022}. After the ``fast radial scan'' phase on Nov-18, the spacecraft began to rapidly retrograde on Nov-19 and Nov-20 (see Figure \ref{fig:E10} (a) for the spacecraft trajectory in the corotating frame). For better comparison, an expanded view is shown in panel (b). The second and third pyramids also show decent capability of capturing the switchback patches, whereas the first pyramid seems to capture a boundary between the patches. Starting from 7:00 on Nov-20, the remaining patches consistently exhibit a high level of Gaussianity across all scales and locations, resulting in indistinct boundaries between them.

\begin{figure*}[ht]
    \includegraphics[width=.90\textwidth]{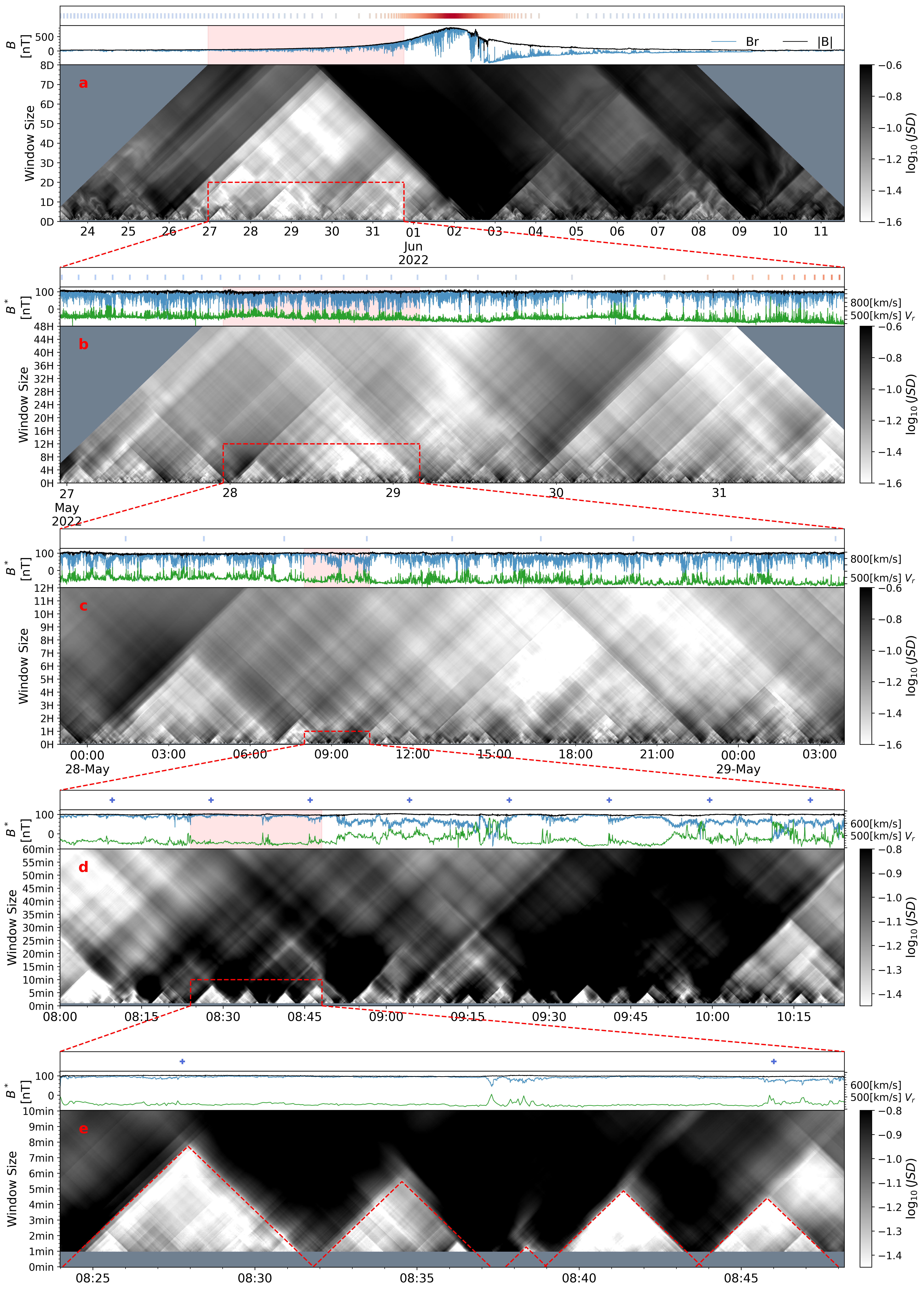}
    \caption{Hierarchic multi-scale Gaussianity Scalogram illustration of mid-latitude coronal hole from PSP encounter 12 inbound. For all subplots, the Carrington longitude of the spacecraft is shown in the top panel. For subplots (a) to (c), the bars are plotted every 1 degree, with colors indicating the heliocentric angular velocity in the solar corotating frame (blue: progradation, red: retrogradation). For subplots d and e, the crosses are plotted every 0.1 degree. The corresponding magnetic field magnitude $B$ and radial component $B_r$ are shown in the second panel of each subplot; and except for subplot (a), the magnetic field is normalized with helio-radial power law fit. The radial solar wind speed $V_r$ is also shown in subplot (b) to (e).}
    \label{fig:multi scale}
\end{figure*}

Figure \ref{fig:multi scale} presents a hierarchic GS of the mid-latitude coronal hole from the inbound of PSP E12. In panels (d) and (e), focusing on the smallest scales resolvable by the JSD (approximately 1 minute, corresponding to around 20,000 data points for the shortest interval. For details on how the number of data points influences this analysis, see Appendix), we observe a surprising number of structures with distinct boundaries. In fact, these structures, typically lasting 1-10 minutes, are omnipresent in the Alfv\'enic solar wind for all PSP encounters. Notably, they are not limited to winds with a clear coronal hole origin, such as those in the outbound paths of E12 (for more details, see the video in the supplementary materials). These structures are typically separated (interrupted) by radial jets (i.e. individual switchbacks), and these separations are frequently accompanied by close to kinetic scale ($\lesssim$ 5 seconds) fluctuations that are bursty and short-lived in all three components of magnetic field. For further illustration, refer to the skewness scalogram video in the supplementary materials. Unlike the spatial structures shown in panels (a), (b), and (c) (as well as in Figure \ref{fig:E10} and Figure \ref{fig:ulysses}), the longitude change of the spacecraft for each structure in panel (d) and (e) is less than 0.1 degree, as indicated by the crosses plotted every 0.1 Carrington longitude in the top bar. Therefore, these structures are likely temporal, i.e. advected by the solar wind. All of these features are highly compatible with the ``jetlets'' observed in equatorial coronal holes \citep{raouafi_magnetic_2023}, and therefore could potentially be the ``building blocks'' of the solar wind. In fact, even finer structures can be found with the normalized standard deviation ($\sigma_{B^*}/\langle B^* \rangle$) scalogram and skewness scalogram shown in Figure \ref{fig:skewness}. For example, the small white pyramid around 8:36 in Figure \ref{fig:multi scale}e has two 30-seconds long substructures nested beneath in Figure \ref{fig:skewness}. These seconds-long structures are intervals with smaller standard deviation compared to the surroundings, and their interruptions are temporally compatible with the ``picoflare'' \citep{chitta_picoflare_2023}.

\section{Discussion}

\begin{figure*}[ht]
    \includegraphics[width=.70\textwidth]{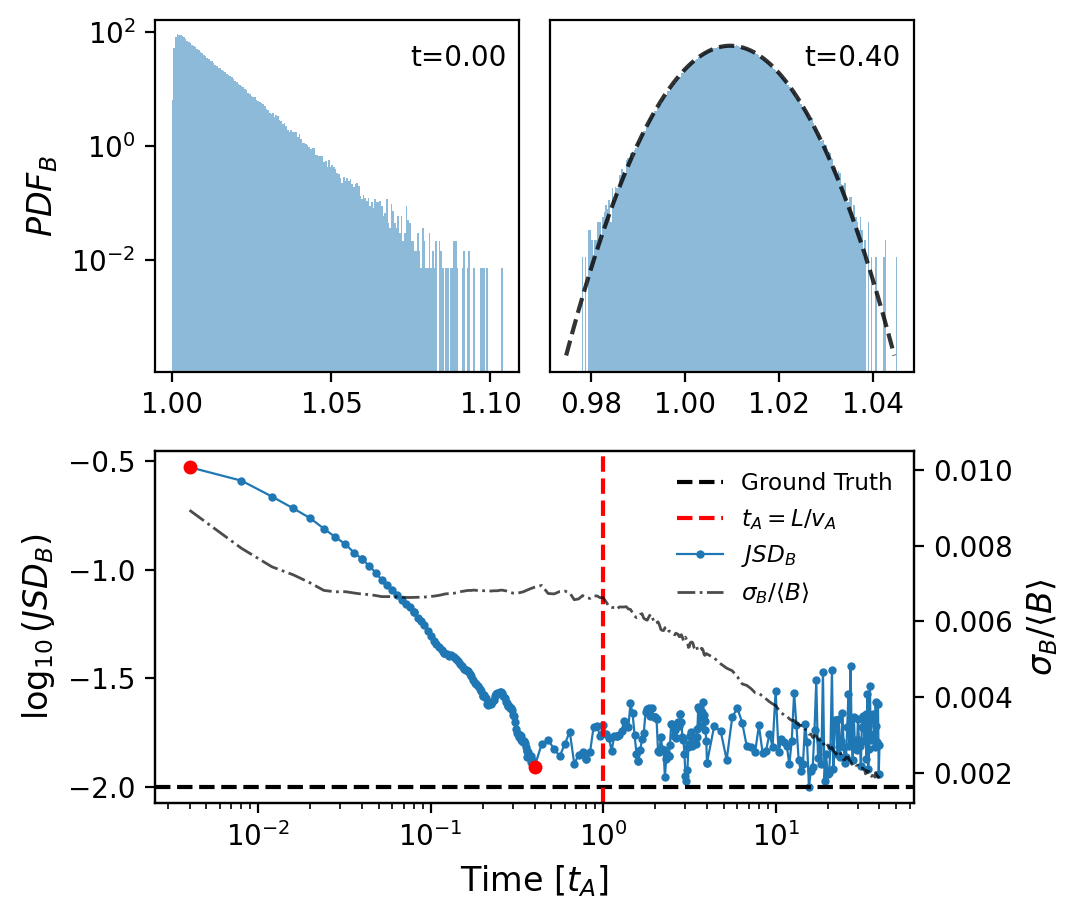}
    \caption{Relaxation of magnetic magnitude $B$ in Alfv\'enic MHD turbulence simulation. Upper panels: (Left) Probability distribution of $B$ ($PDF_{B}$) at $t=0.00\ t_A$, where $t_A=L/v_A$ is the Alfv\'en time and $L$ is the size of simulation box, $v_A$ is the Alfv\'en speed; (Right) $PDF_{B}$ at $t=0.40\ t_A$. Lower panel: Time evolution of the Jensen-Shannon Distance between $PDF_{B}$ and Gaussian Distribution (blue line), and the normalized standard deviation of $B$ (dash dotted line). The time axis is normalized with the Alfv\'en time $t_A$. The simulation time step of the upper left and right panels are highlighted with two red circles in the lower panel. }
    \label{fig:simulation}
\end{figure*}

These observations indicate that the Alfv\'enic solar wind is permeated with highly Gaussian magnetic magnitude intervals that are often interrupted by radial jets (switchbacks) every 1-10 minutes. In addition, the magnetic fluctuations inside the intervals often resemble the small amplitude outward propagating linear Alfv\'en waves. It is therefore reasonable to model the system using small amplitude Alfv\'enic MHD turbulence. Figure \ref{fig:simulation} shows the temporal evolution of the $JSD(PDF_{B}, \mathcal{N})$ of a 3D MHD small amplitude Alfv\'enic turbulence simulation \citep{shi_evolution_2023}. The simulation is run with $512^3$ periodic box, and is initialized with unidirectional small amplitude linearly polarized Alfv\'en waves with isotropic wave vector spectrum (see Methods section for more details). At $t=0.00\ t_{A}$ (Alfv\'en crossing time $t_A = L/v_A$, where $L$ is the simulation box size), $PDF_{B}$ deviates significantly from a Gaussian distribution due to the small amplitude shear Alfv\'en wave initialization (fluctuations in $B$ are positive definite). The corresponding JSD is highlighted as the first red dot in the lower panel and is much larger than 0. Surprisingly, within one Alfv\'en crossing time at $t=0.40\ t_A$, the distribution of $B$ rapidly relaxes to a near-perfect Gaussian distribution, and the JSD rapidly drops towards the ground truth value (see Benchmark in Methods). As the simulation evolves, the JSD remains considerably small and thus the distribution of $B$ remains very close to Gaussian. The simulation indicates that Gaussian is the natural relaxation state for magnetic magnitude in small amplitude Alfv\'enic turbulence, consistent with the ubiquitous 1-10 minutes Gaussian intervals found in the solar wind. However, a 3D analytical model of a switchback with constant magnetic magnitude and fully open field lines is not yet available (see \cite{tenerani_magnetic_2020, squire_construction_2022, shi_analytic_2024,matteini_alfvenic_2024} for recent progress on switchback modeling). Therefore, our simulation can not reproduce the realistic physical condition of the solar wind turbulence in which large amplitude spherically polarized Alfv\'enic fluctuations dominate.

Nevertheless, the simulation suggests that information is fully exchanged within the system, as it propagates at Alfv\'en speed throughout the simulation box. This allows $B$ to relax to a Gaussian distribution, which occurs within about 0.5 $t_A$, i.e. the time it takes for Alfv\'en waves to carry information from the center of the simulation box to its edges. \textcolor{red}{This is indicative of a dynamic Gaussianization of $B$ in MHD turbulence, which requires MHD waves to facilitate this process due to causality in this physical system. However, for the hours and days long interval, the physical distance PSP and Ulysses trajectories covered are often degrees or tens of degrees apart in Carrington longitude (and latitude). Due to the lack of perpendicular information carrier, plasma that is degrees of Carrington longitude apart should not be considered as part of the same turbulent plasma parcel, and thus our simulation does not apply to such situation.}

Therefore, alternative explanations are necessary for the observed hour-long (and longer) Gaussian structures. The simplest explanation for the Gaussian $B$ structures originating from coronal holes (mid-latitude coronal holes from E10 and E12, and polar coronal holes from Ulysses) is the pressure balance between the open coronal field lines. Close to the sun, the solar wind originating from the coronal holes is mostly magnetic dominant (plasma $\beta=2\mu_0 P/B^2 \ll 1$, see e.g. \citep{kasper_parker_2021}). To maintain pressure balance, the open field lines from the same coronal hole tend to evolve to a state in which the magnetic pressure $P_B=B^2/2\mu_0$ is mostly uniform for a given cross section of the magnetic flux tube. In Figure \ref{fig:E10}, the helio-radial power law normalization of $B$ essentially maps the magnetic field line density, which is effectively the magnetic flux density due to the spherical polarization of the Alfv\'en waves, from various radial distances and transverse locations to a single cross-section of the flux tube (for more details of spherical polarization of Alfv\'en waves, see the appendix and \cite{matteini_dependence_2014,matteini_ion_2015}). As a support of this idea, from the PSP observations of E10 and E12 (Figure \ref{fig:E10} and Figure \ref{fig:E12}), the helio-radial power law normalization of $B$ effectively collapses the histogram of $B$ into a delta-function-like histogram of $B^*$. This is indicative of identical field line density within a single coronal hole due to the magnetic pressure balance. The detailed distribution of $B^*$ is hence the feature of the noise in magnetic magnitude within a single coronal hole, which can be considered as a one-dimensional random walk (continuous addition of small amplitude random fluctuations that can be considered as samples drawn from the same stochastic source throughout its passage from the base of the corona to the spacecraft). Therefore, the Gaussian distribution of $B^*$ can be easily explained as the result of the stopped random walk according to central limit theorem. Nevertheless, difficulties remain for the physical origin of the hour-long structures. They may be the manifestation of the denser field line density originating from a single supergranule based on its connection with switchback patches, but a more detailed discussion lies beyond the scope of this study.

Finally, the existence of a stable power law dependence of $B$ with regard to heliocentric distance $R$ itself already sheds light on the physics of the solar wind originating from coronal holes. As solar activities ramp up for solar cycle 25, 4 out of the 5 recent encounters (E10, E11, E12, E14) of PSP show systematic preference for a single helio-radial power law index, which consistently deviates from $R^{-2}$. However, the $R^{-2}$ power law, expected only from the dominant radial component $B_r$ as a result of the Parker Spiral (conservation of magnetic flux in spherical expansion), is not strictly applicable to $B$, especially for PSP, due to the ubiquitous switchbacks. Due to the relation between $B$ and the local magnetic flux density, this is indicative of a stable expansion rate for the magnetic flux tube in the magnetic dominant wind ($\beta \ll 1$) close to the sun. Such an expansion rate is crucial for the estimation of the WKB evolution of the fluctuation quantities like the magnetic and velocity field \citep{hollweg_alfven_1973,heinemann_non-wkb_1980,velli_waves_1991,huang_conservation_2022}. It should be noted that the fit indices of $B$ coincide with the helio-radial dependence of the electron density compiled from Quasi Thermal Noise \citep{kruparova_quasi-thermal_2023,moncuquet_first_2020}, indicating that the deviation from $R^{-2}$ could be the evidence of active acceleration of the solar wind.

\section{Conclusion and Summary}

Compiled from the almost featureless magnetic magnitude time series from the solar wind, the Gaussianity Scalogram (GS) unveils a striking number of fractal-like magnetic structures spanning across over seven orders of magnitude in time. These structures include spatial structures like polar coronal holes \citep{mccomas_three-dimensional_2003}, mid-latitude coronal holes \citep{badman_prediction_2023}, and switchback patches \citep{bale_interchange_2023}. They also include temporal structures compatible with ``jetlets'' \citep{raouafi_magnetic_2023} and ``picoflare'' \citep{chitta_picoflare_2023}, which are often interrupted by the radial jets (switchbacks). In addition, three-dimensional MHD simulations have shown that Gaussian is the natural relaxation state for small amplitude unidirectional Alfv\'enic turbulence. The minute-long structures are hence likely to be the natural products of Alfv\'enic MHD turbulence. Thus, it is now clear that the Alfv\'enic solar wind is permeated with these intermittent Gaussian $B$ structures, which are self-similarly organized from seconds to years, and are possibly the remnants of the magnetic structures on the solar surface \citep{uritsky_multiscale_2012,aschwanden_self-organized_2011,aschwanden_25_2016,bale_interchange_2023,raouafi_magnetic_2023,chitta_picoflare_2023}. This paper reveals just a fraction of the rich structures uncovered by the GS from the solar wind time series. The GS proves to be a versatile tool, essential not only for deciphering the structure and dynamics of plasma and magnetic fields, one of the key objectives of the PSP mission \citep{fox_solar_2016}, but also for revitalizing decades-old solar wind data from missions like Helios, Ulysses, and WIND. These efforts unveil new physics previously hidden within these data sets. Additionally, the GS applicability may extend beyond solar wind analysis, potentially serving other kinds of high resolution stochastic time series data.

\section*{Acknowledgment}
Z.H. thanks Jiace Sun, Benjamin Chandran, Anna Tenerani, Victor R\'eville for valuable discussions. C.S. is supported by NSF SHINE \#2229566. C.S. acknowledges the support of Extreme Science and Engineering Discovery Environment (XSEDE) EXPANSE at SDSC through allocation No. TG-AST200031, which is funded by National Science Foundation grant number ACI-1548562 \citep{j_towns_xsede_2014}. C.S. also acknowledges the support of Derecho: HPE Cray EX System (https://doi.org/10.5065/qx9a-pg09) of Computational and Information Systems Laboratory (CISL), National Center for Atmospheric Research (NCAR) (allocation No. UCLA0063). This research was funded in part by the FIELDS experiment on the Parker Solar Probe spacecraft, designed and developed under NASA contract UCB \#00010350/NASA NNN06AA01C, and the NASA Parker Solar Probe Observatory Scientist grant NASA NNX15AF34G, and NASA HTMS 80NSSC20K1275. M.V. acknowledges support from ISSI via the J. Geiss fellowship.

\subsection*{Data Availability}
The PSP and Ulysses mission data used in this study are openly available at the NASA Space Physics Data Facility (https://nssdc.gsfc.nasa.gov) and were analyzed using the following open-source softwares: \texttt{NumPy} \citep{harris_array_2020}, \texttt{SciPy} \citep{virtanen_scipy_2020}, \texttt{Matplotlib}, \texttt{SPEDAS} \citep{angelopoulos_space_2019}, \texttt{Numba} \citep{lam_numba_2015}.



\newpage
\appendix

\section{Jensen-Shannon Distance, Gaussianity scalogram and Benchmark}\label{sec:benchmark}

\begin{figure}[ht]
    \includegraphics[width=.98\textwidth]{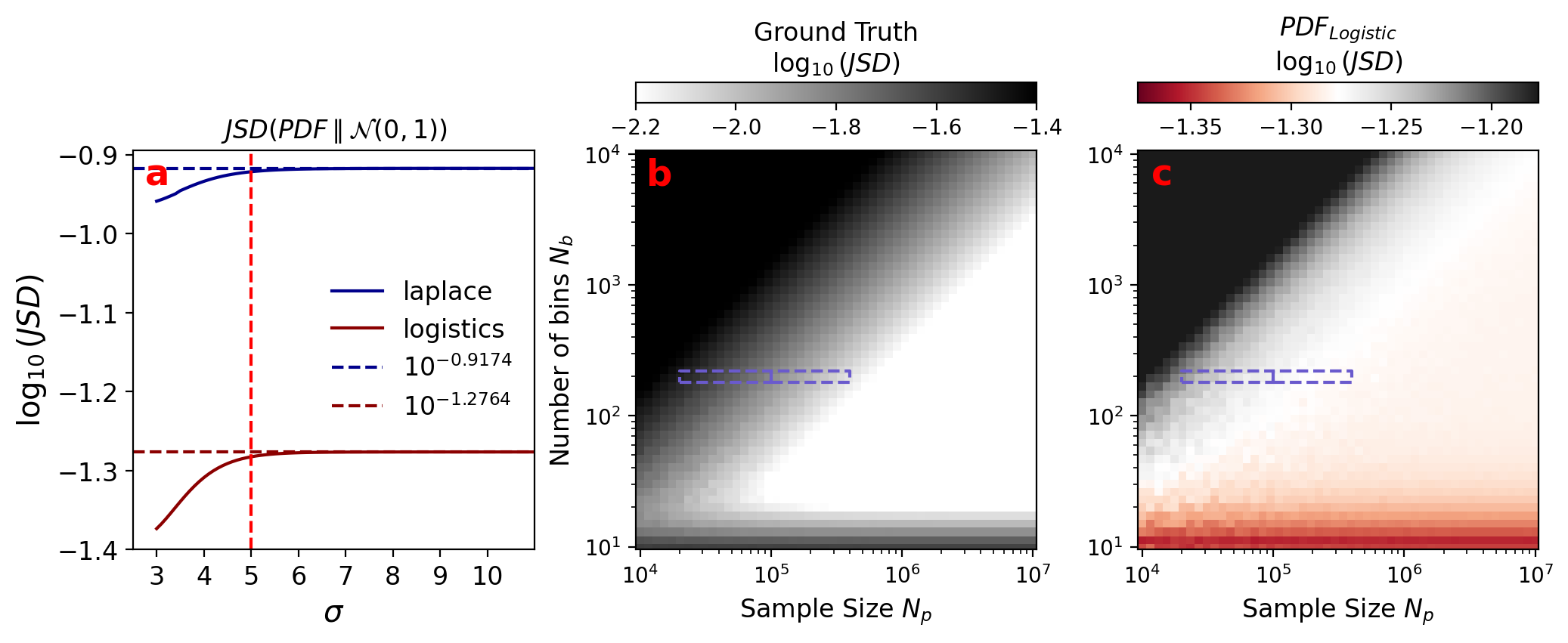}
    \caption{Benchmark of all three free parameters: number of standard deviation $\sigma$, number of bins, and number of points. (a): the Jensen-Shannon distance between a given probability distribution function and the standard Gaussian distribution $\mathcal{N}(0,1)$; (b): the ``Ground Truth'' values, i.e. the JS distance of the histogram with $N_b$ number of bins estimated from $N_p$ number of points from $\mathcal{N}(0,1)$; (c): the same x and y bins as panel (b), with values of $JSD(PDF_{Logistic} \parallel \mathcal{N}(0,1))$.}
    \label{fig:benchmark}
\end{figure}

The Jensen-Shannon Distance is the square root of Jensen-Shannon Divergence \citep{lin_divergence_1991} which is the symmetrized and smoothed version of Kullback–Leibler Divergence \citep{kullback_information_1951}. Due to its symmetry and smoothness, Jensen-Shannon Distance is an ideal metric for the similarity between the observed magnetic magnitude distribution and the Gaussian distribution. For two discrete probability distribution functions $P$ and $Q$ defined in the same space $\mathcal{X}$, the Jensen-Shannon Divergence is calculated as following:
\begin{eqnarray}
    \mathrm{JSD}(P \| Q)=\frac{1}{2} D_{\mathrm{KL}}(P \| M)+\frac{1}{2} D_{\mathrm{KL}}(Q \| M)
\end{eqnarray}
where $M=(P+Q)/2$ is the mixture distribution of $P$ and $Q$, and $D_{\mathrm{KL}}(P \| Q)$ is the Kullback–Leibler Divergence:
\begin{eqnarray}
    D_{\mathrm{KL}}(P \| Q)=\sum_{x \in \mathcal{X}} P(x) \log \left(\frac{P(x)}{Q(x)}\right)
\end{eqnarray}
In this study, we use \texttt{scipy.spatial.distance.jensenshannon} \citep{virtanen_scipy_2020} to calculate the Jensen-Shannon Distance. This program uses natural base logarithm in Kullback–Leibler Divergence, and therefore the final Jensen-Shannon distance is bounded by $[0,\sqrt{ln(2)}]$. \textcolor{red}{Note that kurtosis has been commonly used in past studies to quantify the Gaussianity of probability distribution functions (see e.g., \cite{sioulas_magnetic_2022}). However, kurtosis is only a necessary condition for Gaussianity; for instance, a uniform distribution can have a kurtosis of 3. Therefore, in this study, we chose JSD for its simplicity and interpretability in quantifying Gaussianity. Nonetheless, recent studies have shown that Wasserstein distance \citep{xia_efficient_2024,xia_squared_2024} has advantages over divergence methods, and we plan to investigate these methods in future work.}

The Gaussianity scalogram (GS) is a map where the vertical axis is window size ($win$) and the horizontal axis is the central time of each interval ($t_{mid}$), together forming a scalogram of Jensen-Shannon distance between the normalized probability density function of a given interval $PDF(t_{mid}, win)$ and the standard Gaussian distribution $\mathcal{N}(0,1)$, i.e. $JSD(PDF(t_{mid}, win), \mathcal{N}(0,1))$, or simply $JSD(PDF, \mathcal{N})$. To calculate $PDF(t_{mid}, win)$ from the ensemble of samples from a given interval, there are three controlling parameters: sample size $N_p$, number of bins $N_b$, and number of standard deviation considered $\sigma$. In addition, for benchmark purposes, it is necessary to calculate the JSD between some well-known symmetric distributions and standard Gaussian distribution. The summary of the influence of the controlling parameters and the comparison with well-known distributions are shown in Figure \ref{fig:benchmark}.  

The JSD between Laplace and Logistic distributions and Gaussian distribution as a function of standard deviation range considered is shown in panel (a). The JSD value stablizes approximately at 5 $\sigma$, and therefore for all GS shown in this paper, the $PDF$ are all compiled for $\pm 5 \sigma$. To see how $N_p$ and $N_b$ control the JSD value, samples are repeated drawn from a true Logistic distribution to calculate $JSD(Logistic,\mathcal{N}|N_p, N_b)$. In panel (c), we see a much stablized region for large enough $N_p$ and not-too-large $N_b$ (The stable region is orange-ish because theoretical value at 5 $\sigma$ is slightly smaller than the true value shown in panel (a) as dark red horizontal dashed line). Two purple dashed regions are highlighted in panel (c), where the right one indicates the parameter space used for low resolution GS shown in Figure \ref{fig:E10} and \ref{fig:E12}, and the left one corresponds to the high resolution version shown in Figure \ref{fig:multi scale} (c-e). 

In addition, $N_p$ and $N_b$ also influence the ground truth value, i.e. the Jensen-Shannon distance between an ensemble statistically drawn from Gaussian generator and the real Gaussian PDF, which is not available in closed form \citep{nielsen_jensenshannon_2019}. To obtain the ground truth value, the PDF is a histogram of equally spaced $N_b$ bins located within $\pm$ 5 $\sigma$ compiled from $N_p$ independent samples drawn from a standard Gaussian source \texttt{numpy.random.randn} \citep{harris_array_2020}, and then the JSD is the averaged distance between the statistically calculated PDF($N_p,N_b$) (repeated 30 times for each $N_p$ and $N_b$) and the true Gaussian PDF. The standard deviation is found to be small for a given tuple of $N_p$ and $N_b$. The resulting $N_p$-$N_b$ map is shown in panel (b), and the two parameter space considered are also shown as purple dashed regions. Even for the poorest case ($N_p\sim 20000$), the ground truth value is still sufficiently away from JSD($Logistic,\mathcal{N}$). 

\textcolor{red}{Finally, it should be pointed out that slightly similar multi-resolution visualization (scalogram) methods based on Jensen-Shannon divergence have been proposed in other fields. For example, \cite{torres_multiresolution_2007} use JSD to measure the divergence between wavelet coefficients of speech signals in consecutive time windows, focusing on enhancing speech recognition in noisy environments. \cite{zhang_different_2017} apply JSD to compare the probability distributions of EEG segments during awake and NREM sleep states to analyze EEG complexity. Both studies utilize JSD to compare distributions within their respective datasets, not against a Gaussian distribution. In contrast, our work applies JSD specifically to measure how closely the normalized magnetic magnitude data from the solar wind aligns with a Gaussian distribution. This unique approach allows us to identify and visualize various solar wind structures, such as coronal holes, across multiple temporal scales using the Gaussianity Scalogram. Our domain is solar physics, and by normalizing the magnetic field magnitude with a power law fit, we reveal fractal-like structures in the solar wind spanning seconds to years, offering new insights into their solar origins.}

\section{Spherical Polarization of Alfv\'en Waves}


Although the spherical (arc) polarization of observed Alfv\'en waves is well-known \citep{del_zanna_parametric_2001,riley_properties_1996,vasquez_formation_1996,tsurutani_relationship_1994,wang_large-amplitude_2012,erofeev_characteristics_2019,johnston_properties_2022,hollweg_transverse_1974,squire_construction_2022},  we provide here a description here for spherically polarized Alfv\'en waves in a magnetically dominated plasma (plasma $\beta=2\mu_0 P/B^2\lesssim 0.1$, typical for Alfv\'enic solar wind measured by PSP around perihelion \citep{kasper_parker_2021}).  Similar to \cite{matteini_dependence_2014}, this consider the background magnetic field $\vec B_0$ to have the same constant magnetic magnitude $B$, differently from \cite{hollweg_transverse_1974} where $\vec B_0$ is calculated as the spatial average $\langle \vec B\rangle$, yielding a smaller field magnitude than the radius of the sphere $B$.

For a fluctuation-free magnetic flux tube originating from a coronal hole, the magnetic field is pointing mostly radially in the high corona and solar wind close to the sun. The spherically polarized Alfv\'en waves can therefore be considered as a perturbation to this otherwise quiet system. To maintain the constant $B$ state observed in the solar wind, the additive magnetic perturbation has to ``switchback'' on top of the radial background field. This scenario is depicted in Figure 4 in \cite{matteini_dependence_2014}. The constant magnetic magnitude $B$ is shown as the radius of the circle and the static radial field from coronal hole is $\boldsymbol{B}_0$. To maintain the constant $B$ state, the perturbation to the system $\boldsymbol{B}_1$ is restricted to the semi-circle, and the resultant magnetic vector $\boldsymbol{B}=\boldsymbol{B}_0+\boldsymbol{B}_1$ can thus fluctuate on a constant sphere of $B$. Following this setup, the magnetically dominant ($p\ll B^2/2\mu_0$) incompressible MHD equations can be rewritten as follows ($\rho=const$, $p=const$, $B^2=const$, $\vec B=\vec B_0+\vec B_1$):
\begin{eqnarray}
    \frac{\partial \vec u}{\partial t}=\vec b \cdot\nabla \vec b - \vec u \cdot \nabla \vec u\\
    \frac{\partial \vec b}{\partial t} = \vec b\cdot \nabla \vec u - \vec u \cdot \nabla \vec b
\end{eqnarray}
where $\vec b=\vec B/\sqrt{\mu_0 \rho}=\vec B_0/\sqrt{\mu_0 \rho}+\vec B_1/\sqrt{\mu_0 \rho}=\vec b_0+\vec b_1$. Assuming the frame is co-moving with the bulk flow and the perturbations are Alfv\'enic, i.e. $\vec u=\vec u_1$ and $\vec u_1=\pm \vec b_1$, the equations can be further reduced into a wave equation:
\begin{eqnarray}
    \frac{\partial^2 \vec b_1}{\partial t^2}=(\vec v_a\cdot \nabla)^2\vec b_1
\end{eqnarray}
where $\vec v_a=\vec b_0=\vec B_0/\sqrt{\mu_0 \rho}$. This equation is identical to the circularly polarized Alfv\'en wave equation, except that $\vec B_1$ can be large but restricted to the sphere defined by $B_0$ and the Alfv\'en phase velocity $\vec v_{a}$ is precisely defined (not defined with time-averaged field).

This model leads to some important implications: 1. The spherically polarized Alfv\'en wave is an exact solution and is mathematically identical to the small amplitude shear Alfv\'en mode; 2. If a radial jet is present in the system, i.e. $\vec u_{1r} \parallel \vec B_0$, in accordance with the observed ``switchbacks'', the spherically polarized Alf\'ven waves can only be outward-propagating. This is because to maintain the constant $B$ state, the only possible polarization is $\vec u_1 = -\vec{B}_1/\sqrt{\mu_0 \rho}$; 3. There exists a well-defined background field $\vec B_0$ for the constant $B$ state, and hence the constant magnetic magnitude $B$ can be regarded as a good proxy for the local $\vec B_0$, i.e. the local magnetic flux density. 

In fact, the reversal of the magnetic field line (switchback) does not increase the number of field lines (thus field line density) and the Alfv\'en wave, being a solenoidal mode, does not change the local magnetic flux density. This establishes a connection between the magnetic magnitude (magnetic field line density) and the local magnetic flux density within the magnetically dominated coronal holes close to the sun. The helio-radial normalization of $B$ in the main text can therefore be regarded as mapping the magnetic flux density measured at different radial distances and longitudinal locations back to a cross section of the magnetic flux tube originating from the coronal hole.


\section{PSP and Ulysses Data Analysis}
The Gaussianity scalograms in this paper are compiled from magnetic magnitude time series of PSP and Ulysses. The fluxgate magnetometer of PSP \citep{bale_fields_2016,bowen_merged_2020} offers two versions of level-2 data in RTN coordinates: mag\_rtn\_4\_per\_cyc and mag\_rtn. The GS for intervals longer than one day are compiled with the low resolution (4 samples per 0.874 second) data product and the rest are compiled with the high resolution (256 samples per 0.874 second) mag\_rtn. All magnetic magnitude data points for each interval are treated as independent samples drawn from a stochastic source and therefore the invalid (NaN) values are discarded and no interpolation is applied. The Ulysses magnetic field data is treated the same way.

\section{Three-Dimensional MHD Alfv\'enic Turbulence Simulation}
The simulation is conducted using a 3D Fourier-transform based pseudo-spectral MHD code \citep{shi_propagation_2020,shi_laps_2024}. MHD equation set in conservation form is evolved with a third-order Runge-Kutta method. A detailed description of the simulation set-up and normalization units can be found in \citep{shi_evolution_2023}. Here we briefly summarize the key parameters.

The domain of the simulation is a rectangular box with the length of each side being $L=5$. The number of grid points along each dimension is $512$. To ensure numerical stability, explicit resistivity and viscosity $\eta = \nu=2 \times 10^{-5}$ are adopted besides a de-aliasing.

For the initial configuration, uniform density, magnetic field and pressure are added: $\rho_0 = B = 1$, $P_0 = 0.1006$. The magnetic field has a small angle ($8.1^\circ$) with respect to $x$-axis, and it is inside $x-y$ plane. On top of the background fields, we add correlated velocity and magnetic field fluctuations, i.e. the fluctuations are Alfv\'enic, with 3D isotropic power spectra. The reduced 1D spectra roughly follow $|k|^{-1.3}$.
The strength of the fluctuations is $b_{rms}/B \approx 0.14$ where $b_{rms}$ is the root-mean-square of the magnetic field fluctuation. 

\section{Fluxgate Magnotometer Noise and Zeros-Drift}

\begin{figure}[ht]
    \includegraphics[width=.95\textwidth]{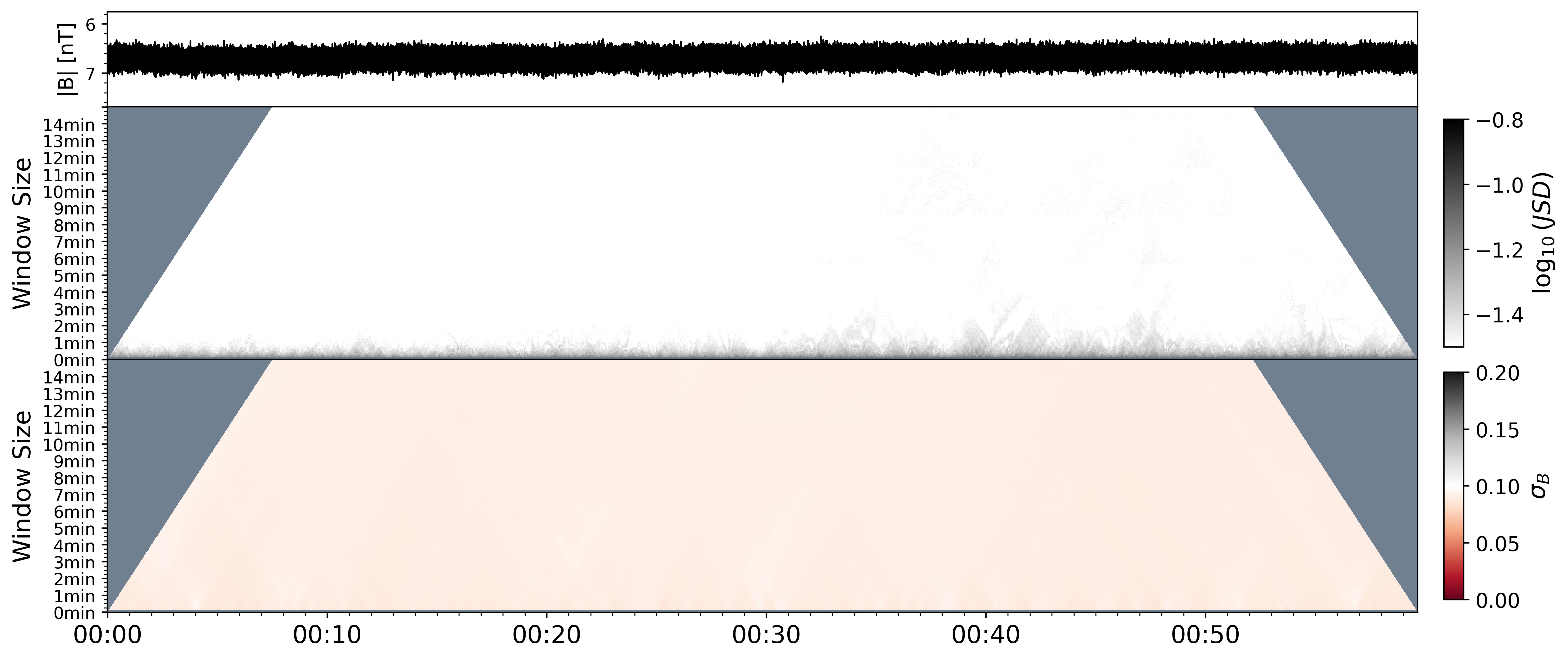}
    \caption{Gaussianity Scalogram of magnetic magnitude from the fluxgate magnotometer noise. From top to bottom: magnetic magnitude $B$ timeseries; GS of $B$; standard deviation scalogram of $B$.}
    \label{fig:noise benchmark}
\end{figure}

There are several sources of error in the PSP fluxgate magnetometer measurements \cite{bale_fields_2016}, including the instrumental noise as well as uncertainty in the zero offsets which drift in time \cite{bowen_merged_2020}. The instrumental noise of each vector component is approximated as Gaussian white noise with a standard deviation $\sigma\simeq0.05 nT$, and together produce a noise with a standard deviation of $\sigma_{noise}\sim 0.1 nT$ for the magnetic magnitude. $\sigma_{noise}$ is usually much smaller than the standard deviation of the {\it in situ} measured $\sigma_B$ for all scales that we are interested in. Nevertheless, the GS of a ground measured one-hour magnetic field time series for calibration is shown in Figure \ref{fig:noise benchmark}. The distribution of the noise signal is universally Gaussian regardless of scales and location, and the standard deviations are unanimously small. Therefore, most of the Gaussian structures we show in the paper are real signals rather than instrument noise.

The error from drifting spacecraft offsets is a significantly larger contribution to the error as the approximated zero-offsets drift over time and are calibrated each day \cite{bowen_merged_2020}. The drift of the spacecraft offsets, which is thought to occur due to slowly varying currents on the spacecraft is not well constrained and varies over time. This error is not Gaussian in nature, but should introduce small offsets in the measured field from the real background magnetic field. Spacecraft rolls are used to determine zero-offsets in both the inbound and outbound phases of each orbit, and are updated daily through optimizing the measurements to ensuring that spherically polarized magnetic field intervals maintain a constant magnitude. Typical offset values drift about $0.5 nT$/day. Due to the continuous drift and non-Gaussian nature, the sub-day ($\lesssim 5 Hr$) structures are not strongly affected by the zeros-drift. And the days-long structures are also not affected because of the instrument calibration of the zeros-offset.


\section{Supplementary Materials}

\subsection{Supplementary Videos}
This manuscript contains one supplementary video: \texttt{gaussianity\_skewness\_scalogram\_E12.mp4}. 

It shows the Gaussianity scalogram and normalized standard deviation scalogram of magnetic magnitude with window sized from 30 second to 15 minutes for the whole Parker Solar Probe E12. This video aims to show the self-similar magnetic structures revealed by GS and the corresponding sub-structures from the normalized standard deviation scalogram. For the first one minute, the GS looks different because of the low sampling rate of the fluxgate magnetometer. It also shows the Skewness scalogram and normalized standard deviation scalogram of magnetic magnitude with window sized from 1 second to 5 minutes for the whole Parker Solar Probe E12. This video aims to show the systematic tendency for magnetic holes in the magnetic magnitude distributions.

\clearpage
\subsection{Supplementary Figures}

\begin{figure*}[ht]
    \includegraphics[width=.95\textwidth]{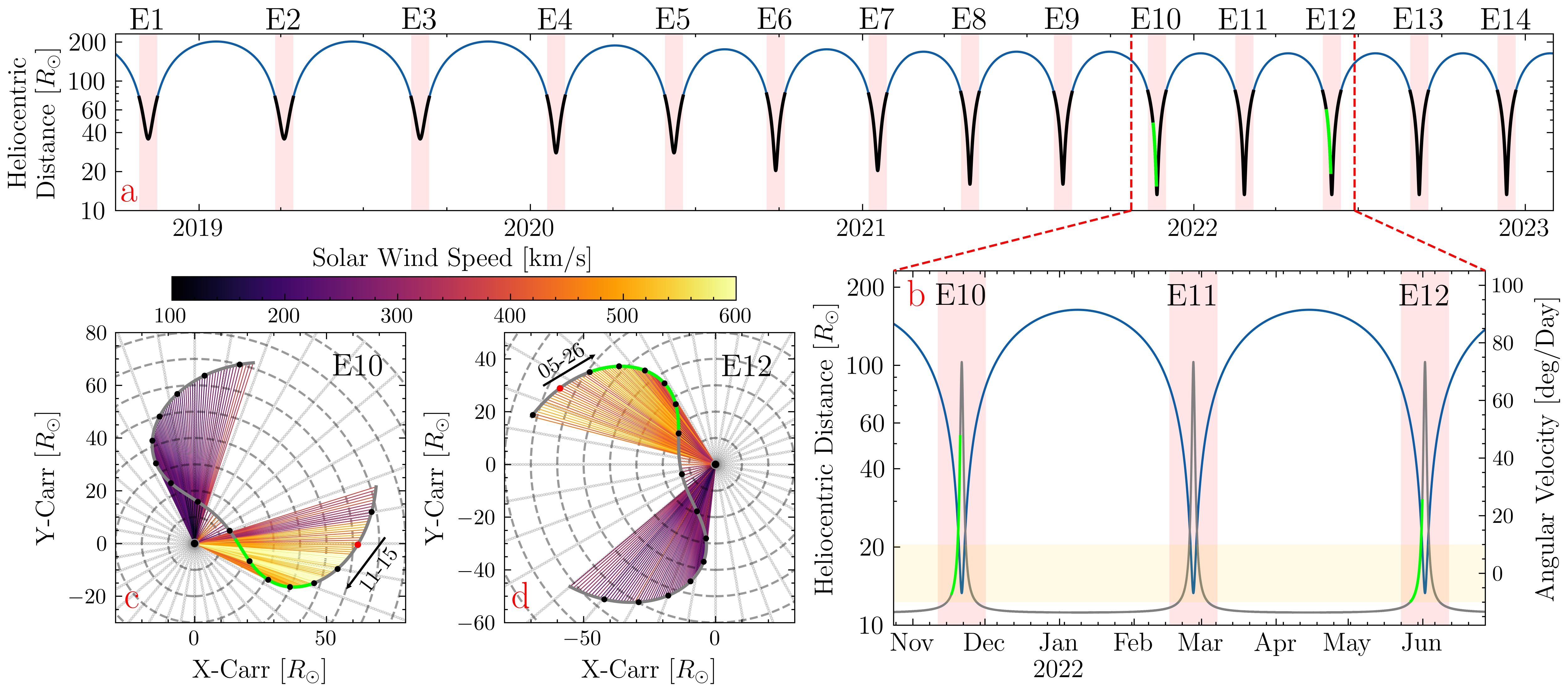}
    \caption{Panoramic plot of the data considered in this study. (a): This panel shows the heliocentric distance of the spacecraft from encounter 1 to 14. The data analyzed in this study is from $\pm$ 10 days around the perihelia, which are highlighted with black lines and pink shaded areas. The two normally distributed long intervals under investigation are represented by the two green segments. (b): This panel provides a detailed illustration of E10 to E12, with the spacecraft's angular velocity in the Carrington corotation frame displayed on the twin axis. The corotating periods ($\omega < 10\ [deg/Day]$) are marked with golden shaded areas, and the selected intervals are highlighted in green on top of the angular velocity profile. (c) and (d): These panels provide a synopsis plot of E10 and E12 spacecraft trajectories from $\pm$ 8 days around the perihelion in the Carrington corotating frame. The starts of each day are indicated by black dots, and the two arrows show the spacecraft's entering directions, with the corresponding dates highlighted by red circles. The solar wind streamlines are colored according to the 10-minute averaged solar wind speed and are plotted every 2 hours. The two selected intervals are also highlighted in green.}
    \label{fig:panorama}
\end{figure*}

\begin{figure*}[ht]
    \includegraphics[width=.90\textwidth]{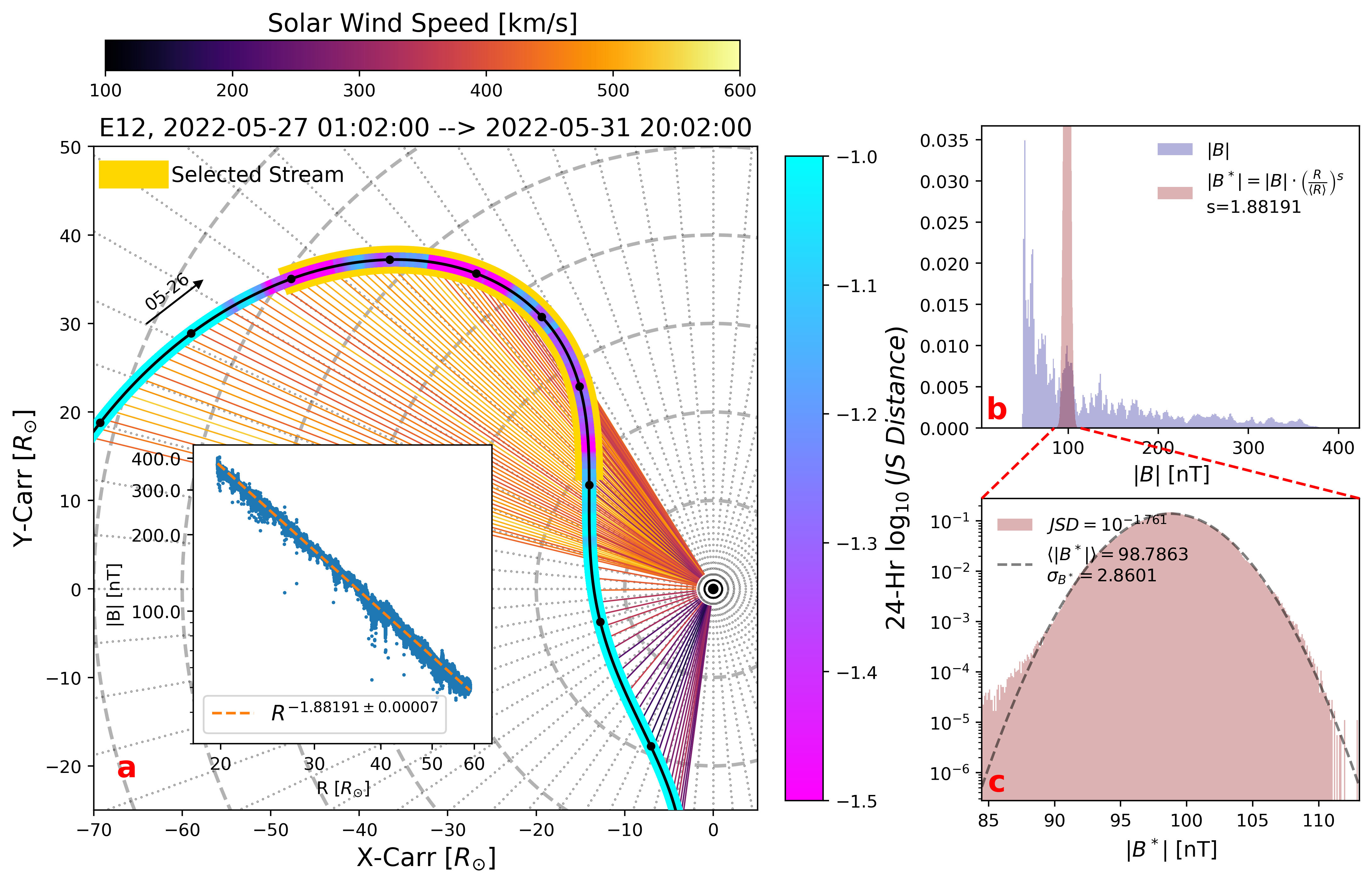}
    \includegraphics[width=.95\textwidth]{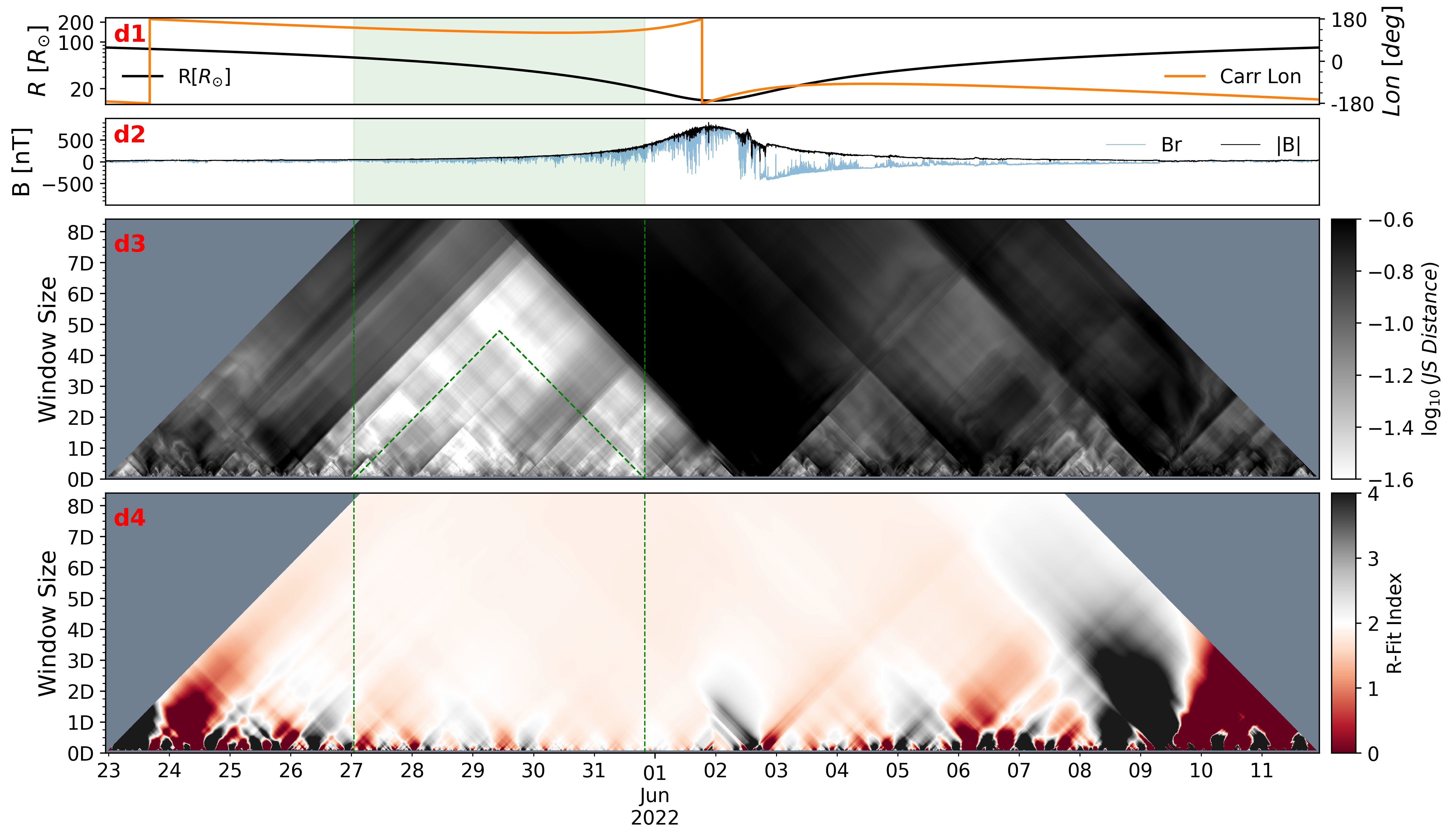}
    \caption{Selected interval from E12.}
    \label{fig:E12}
\end{figure*}

\begin{figure*}[ht]
    \includegraphics[width=.95\textwidth]{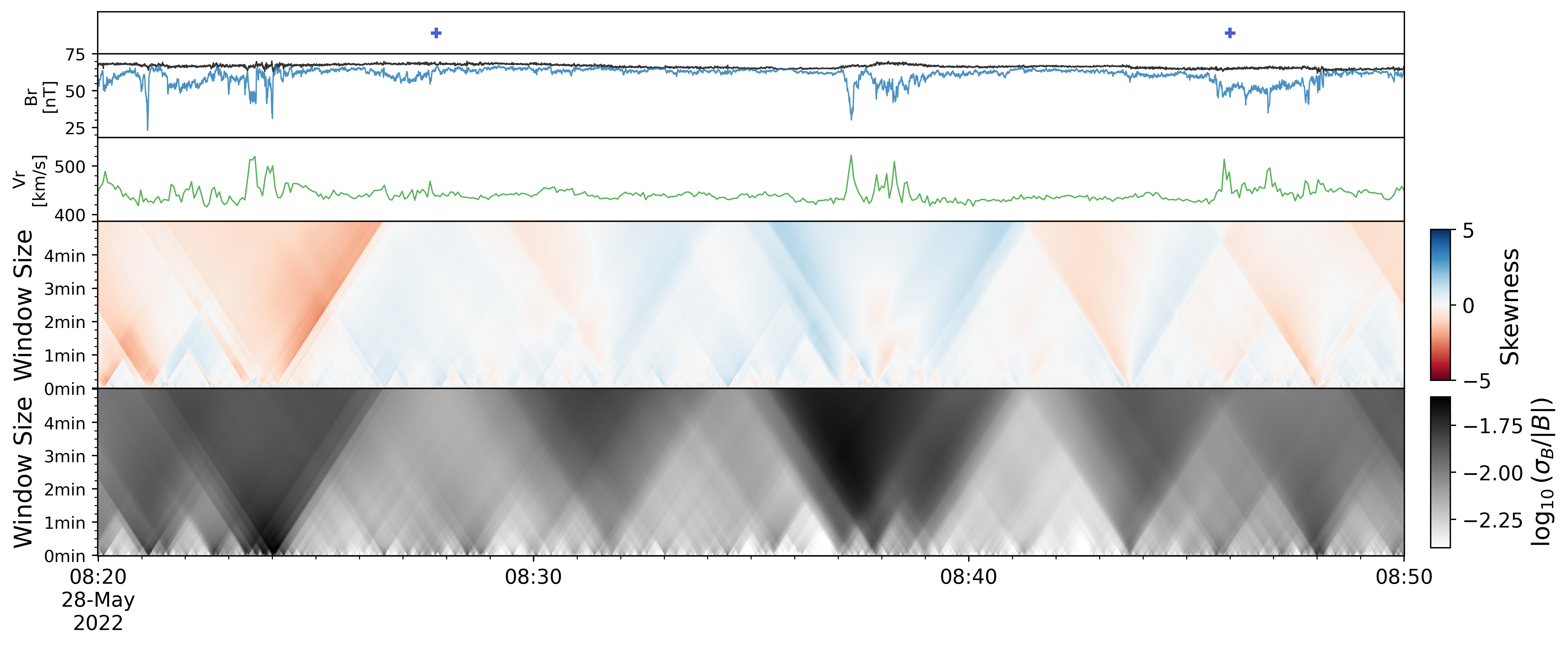}
    \caption{Skewness and normalized standard deviation scalogram. From top to bottom: spacecraft carrington longitude plotted with 0.1 degree cadence; magnetic magnitude ($B$, black) and radial component ($B_r$, blue); radial solar wind speed ($V_r$); skewness scalogram of $B$; normalized standard deviation scalogram of $B$}   
    \label{fig:skewness} 
\end{figure*}

\clearpage
\bibliography{sample631}{}
\bibliographystyle{aasjournal}

\end{CJK*}
\end{document}